\documentclass[10pt,conference,compsocconf]{IEEEtran}
\usepackage{cite}
\usepackage{flushend}
\usepackage{hyperref}
\usepackage[utf8]{inputenc}
\usepackage[T1]{fontenc}
\usepackage{fixltx2e}
\usepackage[pdftex]{graphicx}
\usepackage{longtable}
\usepackage{float}
\usepackage{wrapfig}
\usepackage{soul}
\usepackage{textcomp}
\usepackage{marvosym}
\usepackage{latexsym}
\usepackage{amssymb}
\usepackage{hyperref}
\usepackage{url}
\usepackage{color}
\usepackage{verbatim}
\usepackage{mdwlist}
\usepackage{pifont}
\usepackage[cmex10]{amsmath}
\usepackage[caption=false]{subfig}

\newcommand{\fig}[3]{
  \begin{figure*}[!t]
    \subfloat[Uniform website popularity]{
      \includegraphics[width=3.4in]{./#1.png}
     \label{subfig:#1}
   }
   \subfloat[Power-law website popularity]{
     \includegraphics[width=3.4in]{./#2.png}
     \label{subfig:#2}
   }
  \caption{#3}
   \label{fig:#1}
\end{figure*}}

\title{Beyond the Blacklist: Modeling Malware Spread and the Effect of
  Interventions}

\author{\IEEEauthorblockN{Benjamin Edwards\IEEEauthorrefmark{1},
Tyler Moore\IEEEauthorrefmark{2},
George Stelle\IEEEauthorrefmark{1},
Steven Hofmeyr\IEEEauthorrefmark{3} and
Stephanie Forrest\IEEEauthorrefmark{1}\IEEEauthorrefmark{4}}
\IEEEauthorblockA{\IEEEauthorrefmark{1} Department of Computer Science \\
University of New Mexico\\
Albuquerque, NM, USA\\
\{bedwards,stelleg,forrest\}@cs.unm.edu}
\IEEEauthorblockA{\IEEEauthorrefmark{2} Department of Computer Science \\
Wellesley College\\
Wellesley, MA, USA\\
tmoore@cs.wellesley.edu}
\IEEEauthorblockA{\IEEEauthorrefmark{4} Santa Fe Institute\\
Santa Fe, NM, USA}
\IEEEauthorblockA{\IEEEauthorrefmark{3} Lawrence Berkeley National Laboratory\\
Berkeley, CA, USA\\
shofmeyr@lbl.gov}}

\begin{document}
\maketitle

\begin{abstract}

Malware spread among websites and between websites and clients is an
increasing problem.  Search engines play an important role in
directing users to websites and are a natural control point for
intervening, using mechanisms such as blacklisting.  The paper
presents a simple Markov model of malware spread through large
populations of websites and studies the effect of two interventions
that might be deployed by a search provider: blacklisting infected web
pages by removing them from search results entirely and a
generalization of blacklisting, called depreferencing, in which a
website's ranking is decreased by a fixed percentage each time period
the site remains infected.  We analyze and study the trade-offs
between infection exposure and traffic loss due to false positives
(the cost to a website that is incorrectly blacklisted) for different
interventions.  As expected, we find that interventions are most
effective when websites are slow to remove infections. Surprisingly,
we also find that low infection or recovery rates can increase traffic
loss due to false positives. Our analysis also shows that heavy-tailed
distributions of website popularity, as documented in many studies,
leads to high sample variance of all measured outcomes.  These result
implies that it will be difficult to determine empirically whether
certain website interventions are effective, and it suggests that
theoretical models such as the one described in this paper have an
important role to play in improving web security.
\end{abstract}

\section{Introduction}
\label{sec:intro}

The network worms which caused havoc ten years ago, such as Code Red,
actively spread by `pushing' themselves onto vulnerable systems
through automated scanning. In contrast, a major problem today is
computer infections that propagate via a `pull'-based mechanism.  For
example, in a drive-by download, an attacker infects a victim
computer's web browser without direct
interaction~\cite{Provos07,Provos:Security08}. In this scenario, the
attacker first compromises an otherwise benign web server, injecting
executable code into its web pages, and then waits for users to visit
the infected website and acquire the infection. Because many users
arrive at websites through search, search engines have become a
crucial battleground over the distribution of malware.

Search providers have an incentive to defend against such attacks
because they degrade search results. A typical approach is that taken
by Google, which attempts to detect and blacklist websites that host
malicious content~\cite{GoogleSafe}. Blacklisting can take the form of
displaying a warning message via a client side browser plugin to
discourage users from visiting a website, or outright removal from the
search results. Blacklisting can be used to combat many types of
malicious content, and in a web environment where new attacks are
developed frequently, it is important to have a general approach to
reducing infection. However, because blacklisting can dramatically
reduce visits to websites, search engines are very careful to avoid
false positives (i.e., flagging an uninfected website as
infected). Such caution can delay responses, which in turn may raise
infection rates.

In this paper we devise a concise Markov model to study how web
infections spread through large populations of websites, and explore
how infections might be contained through blacklisting. We also
propose a generalization of blacklisting called {\em depreferencing},
where a search engine reduces a website's ranking in search results in
proportion to the engine's certainty that the website is infected.
Depreferencing can be more tolerant of false positives than a binary
response such as blacklisting, because the scale of the intervention
can be adjusted to specific levels of false positives. Depreferencing
provides a controllable {\em depreferencing parameter}, $\sigma$, that
can be tuned to achieve specific reductions in infections or false
positives. We derive exact analytic expressions that relate the
depreferencing parameter, $\sigma$, to infection rates and traffic
loss due to false positives. We also identify critical points for the
model parameter values that govern the trade-off between infection and
traffic loss.

\begin{figure}[!t]
  \includegraphics[width=0.5\textwidth]{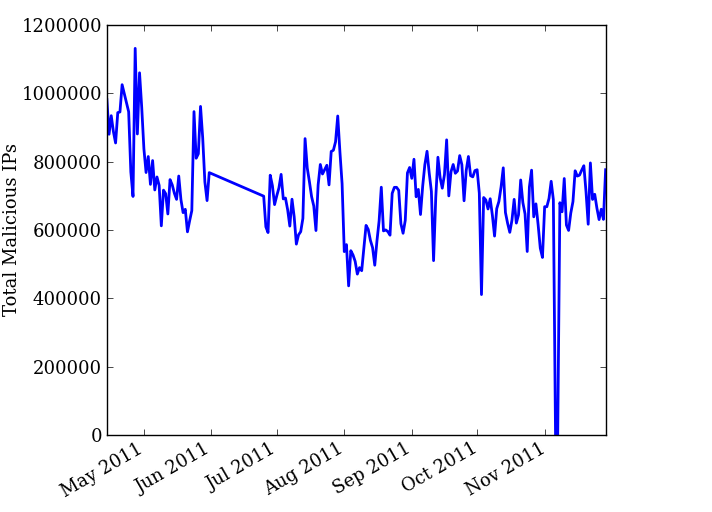}
  \label{fig:ISC}
  \caption{Variation in malicious IP addresses over time.}
\end{figure}

We believe that modeling is particularly well-suited to the task of
examining techniques for controlling malware spread over the web.
First, it allows us to examine unconventional interventions, such as
depreferencing, at low cost. Given the relatively grim status quo in
web security, more radical countermeasures deserve consideration, and
modeling offers a good way to assess the impact of new strategies
without the expense and commitment of an actual
implementation. Second, modeling can deal with the extreme dynamics of
the web better than empirical exploration alone. Our analysis shows
that the heavy-tailed distribution of website popularity leads to high
variance in outcomes. This high sample dependence makes it extremely
difficult to conduct reliable comparative assessments of the benefits
of different interventions, especially with a limited number of
empirical measurements. For example, the infection of a single popular
site can suddenly and dramatically increase overall user infection
rates, an effect that can be seen in Figure~\ref{fig:ISC}, which
contains data on malicious IP addresses collected from the Internet
Storm Center\footnote{\url{http://isc.sans.org}}. We show that this
variance can obscure even large improvements in infection and
recovery. With the modeling approach, we can easily run many
simulations, and more reliably estimate the comparative impacts of
different intervention strategies.

Finally, modeling lets us examine the impact of interventions across
many stakeholders and identify tensions that may arise. For instance,
improved security for search operators and consumers may be achieved
in part at the expense of increased risk of incorrect blacklisting for
website operators. Modeling allows us to more precisely quantify these
trade-offs.

\vspace{-0.1in}

\section{Modeling Infections}
\label{sec:model}

\begin{figure*}[!t]
 \centering
 \begin{minipage}[b]{.5\textwidth}
   \includegraphics[width=\textwidth]{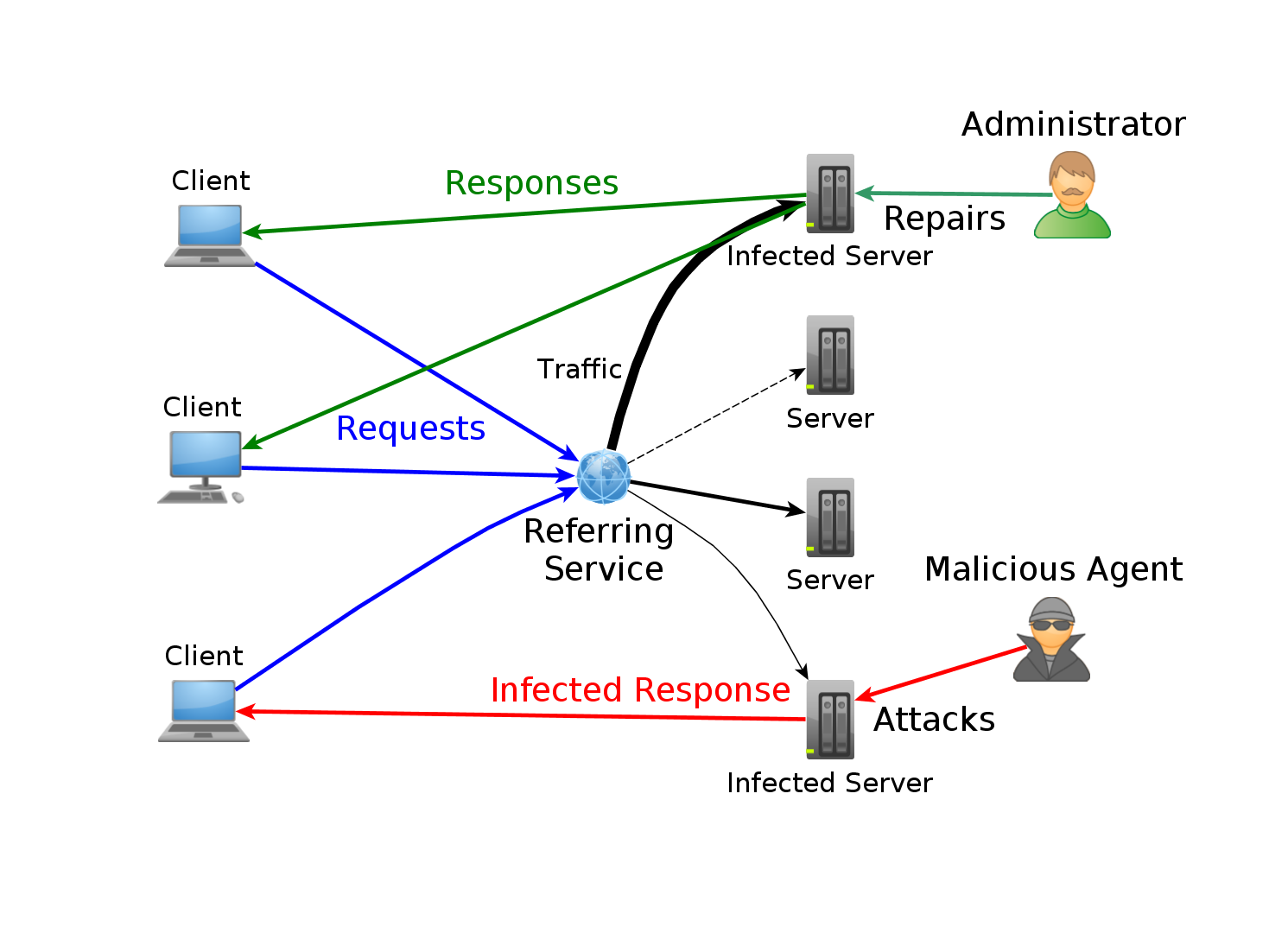}
   \vspace{-0.5in}
   \caption{Server and client infections via search engine referral.}
   \label{fig:RealWorldModel}
 \end{minipage}\hfill
 \begin{minipage}[b]{.5\textwidth}
   \includegraphics[width=\textwidth]{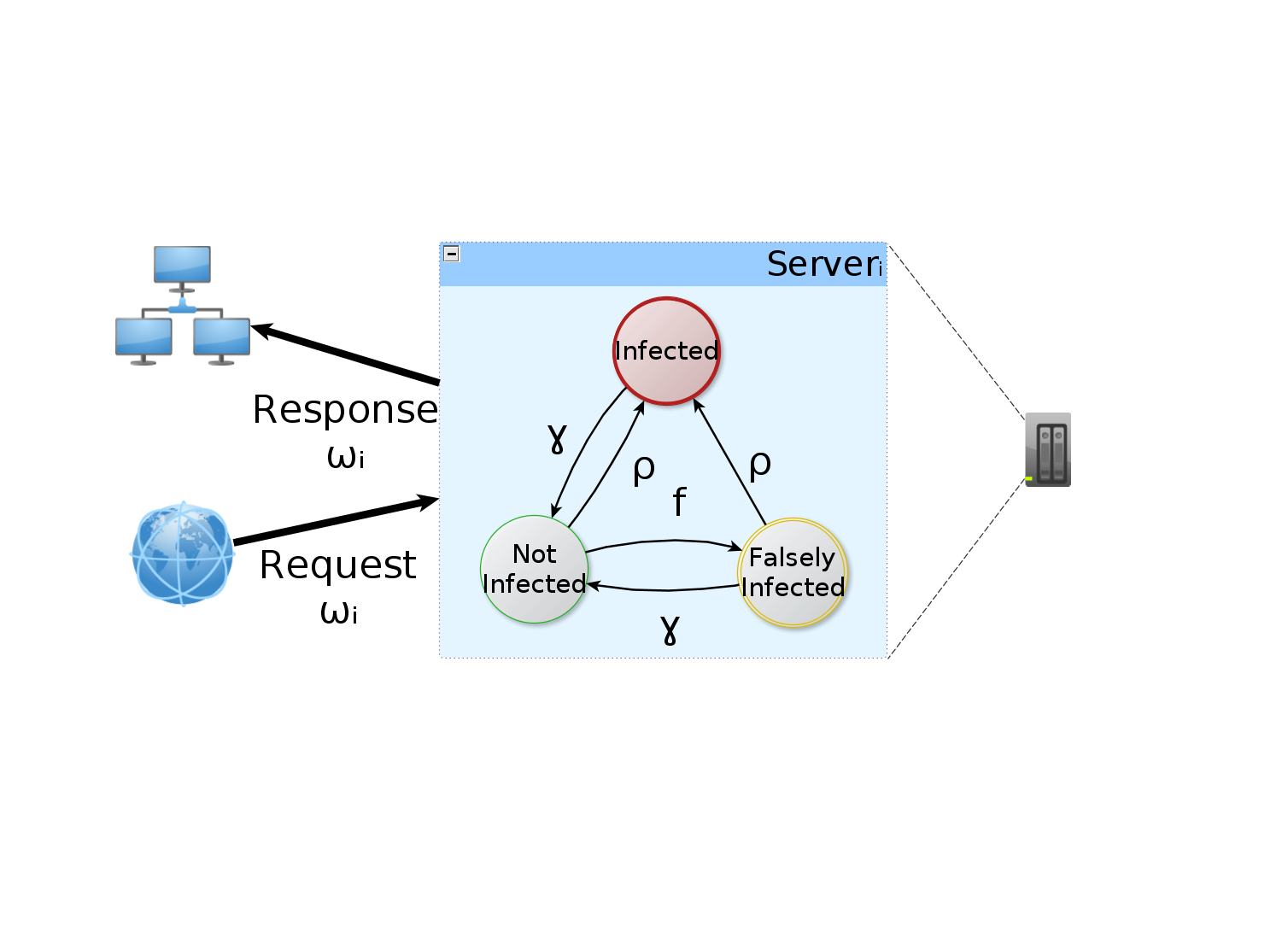}
   \vspace{-0.5in}
   \caption{Model of website infections and client exposure.}
   \label{fig:MarkovModel}
 \end{minipage}\hfill
\end{figure*}

We model a population of servers that is under attack from malicious
agents, as depicted in~\autoref{fig:RealWorldModel}. We do not model
specific types of infections, but rather assume that an infection is
any event that compromises a website such that it could be used to
spread malware to users. Once infected, a server recovers when an
administrator notices the infection and clears it. In this paper we
explore the impact of search provider interventions and so are only
interested in clients that connect to servers via referrals from a
search provider. Hence, in our model, client exposure to infection is
driven solely by website popularity as determined by the search
provider.  In an attempt to improve search results, the search
provider monitors websites to determine whether they are infected, and
may incorrectly identify uninfected websites as infected. We assume
that an administrator clears false identifications of infection at the
same rate as real infections.

Our model includes a population of $n$ websites\footnote{We will use
  the terms website and web server, or simply server,
  interchangeably.}, each with a popularity, $\omega_i$, drawn at
random from a specified distribution. $\omega_i$ represents the total
number of visits a website receives. The key outcome we are interested
in measuring is {\em client exposure}, which is directly proportional
to the expected number of visits that infected websites receive. At
any time, a website is in one of three possible states: infected,
uninfected, or falsely infected (i.e. classified by the search
provider as infected when it is actually not). Each server transitions
between these states at discrete time steps, according to the Markov
chain depicted in~\autoref{fig:MarkovModel}. The key parameters are:
$\rho$, the probability of a website becoming infected; $\gamma$, the
probability of recovering from an infection; and $f$, the probability
of falsely being classified as infected.

We make the simplifying assumption that the probabilities $\rho$,
$\gamma$, and $f$ are constant across the population of servers and
time invariant. Unfortunately, data on the exact distributions of
these parameters are not readily available and often
contradictory. For example, there are no data supporting a systematic
relationship between a website's popularity and its susceptibility to
infection. We examined a sample of websites infected with malware that
were reported to StopBadware from 2007 to 2009, and found that more
popular websites (as ranked by Alexa) are slightly more likely to be
infected. By contrast, Moore et al.~\cite{Moore:CCS11} found that more
popular web search terms are less likely to include infected websites
in their results. In both cases, the effects are small so we argue
that assuming constant probabilities is reasonable.

Our model is discrete time; an alternative approach is to model the
population of servers using differential equations. In the case of
large $n$, the steady state distribution of infection probability
would be exactly the infection rates in a differential equation
model\cite{zou2002}. We use a discrete-time model instead because it
allows us to easily incorporate time-dependent phenomena (such as our
interventions) and distributions of values (such as traffic), and it
is simpler to explore transient effects.

\section{Modeling Interventions}
\label{sec:interventions}

We model two forms of intervention: {\em blacklisting}, which is
currently used by search engines, and a hypothetical approach called
{\em depreferencing}, which offers a way to adjust intervention
parameters to specifically control the trade-off between infections
and traffic loss due to false positives.

\subsection{Blacklisting}

A common approach taken by search engines that detect a compromised
website is to inform the user in the search results (through a client
side application), before the user has a chance to visit the website,
and then to produce further warnings if the user persists in
attempting to visit the website. This is equivalent to blacklisting
the site because users are unlikely to ignore the warning.\footnote{In
  Google, clicking on a result labeled infected takes the user to a
  warning page with a small text URL at the bottom of the page which
  the user has to copy and paste into the browser navigation bar.}
Because blacklisting prevents the website from receiving all or nearly
all of its search traffic, minimizing false positives is
essential. For example, Rajab et al.~\cite{Rajab:TR11} claim that
Google's Safe Browsing infrastructure ``generates negligible false
positives.''

We assume that blacklisting takes a fixed number of time steps to
detect a compromised website and blacklist it. We refer to this as the 
\emph{detection delay}, denoted $\beta$. A website infected at time $t$ will
be blacklisted at time $t+\beta$. Once blacklisted, the traffic to
that the website is set to zero, i.e. $\omega_i=0$. Formally, if a
website, $i$, is infected at time $x$, its traffic, $\hat{\omega}_i$,
at time $t>x$ is

\begin{equation}
\label{eq:blacklisting}
\hat{\omega}_i = 
\begin{cases}
  \omega_i, & \text{if }t-x<\beta \\
  0, & \text{if }t-x \geq \beta
\end{cases}
\end{equation}

The time period $\beta$ captures the notion that it will take a search
engine a certain amount of time to determine that a website is
compromised with high certainty (negligible false positives). Thus
$\beta$ accounts for how frequently the website is crawled, how much
computational time is required to confirm the infection, how much the
search engine is willing to invest in malware detection, and other
possibilities, such as giving the compromised site a certain grace
period to clean up the infection.

In the model, we assume that immediately after a website recovers, its
popularity is restored to its previous value. That is, once a website
has been cleaned, the administrator informs the search engine and the
blacklisting is removed without delay. In reality, there would be a
small delay before the blacklisting is removed. For example, when an
administrator requests Google to run an automated test for malware, it
will take at most a few hours to complete~\cite{GoogleCleanup}, and up
to 24 hours for the malware warning to disappear from all search
results. Because the time period is small and constant, we can exclude
it from our model without significantly changing the results.

\subsection{Depreferencing}

We explore a generalized hypothetical intervention, called {\em
  depreferencing}, which, to the best of our knowledge, is not
actually implemented by any existing search engine. The idea is that
when a search engine detects a possibility of infection in a website,
it reduces the traffic that website receives. This could be
implemented by reducing the rank of that website in the search
results, or probabilistically providing warnings to users. Because the
response does not block all traffic to the website, but rather reduces
the volume of traffic, the detection process can tolerate false
positives, allowing the search engine to react more rapidly and
aggressively. Search providers could use coarser and less precise
detectors to crawl websites more frequently, requiring significantly
less computation time to classify websites as infected.

We model this intervention by reducing the popularity of a website by
a fixed percentage every time step after it is discovered that the
website is infected. If a website is infected at time $x$, an infected
website's traffic at time $t>x$ is

\begin{equation}
\label{eq:depreferencing}
\hat{\omega}_i = 
\begin{cases}
  \omega_i, & \text{if }t-x<\beta \\
  \sigma^{x-\beta+1}\omega_i, & \text{if }t-x \geq \beta
\end{cases}
\end{equation}

\noindent where $0 \leq \sigma \leq 1$ is the {\em depreferencing}
parameter, which controls the strength of the response. Note that
\autoref{eq:blacklisting} is equivalent to \autoref{eq:depreferencing}
when $\sigma=0$. We believe that adjusting search results is a
plausible response that would be easy to implement. For example, a
search engine like Google could simply reduce the page ranks of
infected websites, which should directly affect their popularity in
search results. Similarly to blacklisting, we assume that when a
website recovers from an infection, its popularity is immediately
restored to its original value. Because the response is less drastic,
search engines may be able to reduce the detection delay $\beta$ in
this new intervention.

\autoref{eq:depreferencing} is one of an even more general class of
methods for combating exposure to infection. We could define a general
$g(\omega_i,x)$, such that $g$ is monotonically decreasing in
time. For example $g$ could be a linear or logistic function. We
choose an exponential decline as it seems a natural fit for our
application. Investigation into other forms appropriate for other
applications is left for future work.

As a consequence of the potentially more rapid and hence imprecise
detection of compromised sites, our model includes a constant
probability $f$ that an {\em uninfected} website is classified as
compromised and has its rank reduced. This is in contrast to the
blacklisting approach, where we assume there are negligible false
positives. For depreferencing, we assume that websites that are
incorrectly classified as compromised recover at the same rate,
$\gamma$, as compromised websites. In other words, the process of
recovery is the same whether a website is actually infected or
not. This requires that the administrator realize that the website is
infected (for example, users of Google's Webmaster Tools are notified
when their sites are infected) and that appropriate steps are taken to
correct the problem.

We do not model false negatives, i.e. infected websites that go
undetected, because our model studies the effect of interventions on
client infection rates, and we assume that in both blacklisting and
depreferencing the detection process has similar levels of false
negatives. Hence, the false negative rates should not affect
comparison of the outcomes.  From a practical perspective, data on
false negatives are rare or non-existent because they are extremely
difficult to gather.  We leave the analysis of false negatives to
future work.

\section{Analysis}
\label{sec:analysis}

This section analyzes the mathematical properties of the model
described in the previous section. First we describe the steady state
values of the Markov chain shown in \autoref{fig:MarkovModel}. Second,
we analyze the first and second moments of the random variables that
define the traffic loss and the number of clients exposed to
infection. We then provide expressions that relate the intervention
parameters to the infection exposure and traffic loss, and identify
critical control points.

\subsection{Steady State Distribution}
\label{subsec:steadyState}

Let the state of a server $i$ in the Markov chain in
\autoref{fig:MarkovModel} be the random variable $S_i \in
\{I,N,F\}$, where $I$ denotes infection, $N$ denotes no
infection and $F$ denotes a false positive infection. It is easy to
see that the Markov chain is ergodic except for some degenerate cases
such as $f=1,\gamma=1,\rho=0$. However, such cases are unlikely to
occur in the real world.

Because our Markov chain is ergodic it is guaranteed to converge to a
unique stationary distribution, which is given by
\begin{align}
Pr[S_i=I] &= \frac{\rho}{\rho+\gamma} \\
Pr[S_i=N] &= \frac{\gamma}{(f+\rho + \gamma)} \\
Pr[S_i=F] &= \frac{f \gamma}{(\gamma + \rho)(f + \gamma + \rho)}
\end{align}
  
Moreover, because this is a finite time-homogeneous ergodic Markov
chain, it will have a short mixing time. Hence we focus on the steady-state in
the remainder of the analysis.

\subsection{Client Exposure and Website Loss}
\label{subsec:ExposureLoss}

The probability that a website becomes infected at a time $t-x$ and
remains infected until time $t$ depends on the probability that the
website was not infected at time $t-(x+1)$, became infected at time
$t-x$, and remained infected for the next $x$ timesteps. More
formally, let $I_x$ denote the event that a server $i$ has been in a
state of infection for \emph{exactly} $x$ time steps. Then
\begin{equation}
  \label{eq:InfectionExact}
  Pr[S_i=I_x] =\rho (1-Pr[S_i=I]) (1-\gamma)^x
\end{equation}

Observe that the events $S_i=I_x$ and $S_i=I_{x'}$, with $x \neq x'$,
are mutually exclusive, e.g. a server cannot be infected for exactly 5
and exactly 6 time steps.

Next we derive an expression for the random variable $X_i(\beta,\sigma)$, which
describes the number of clients exposed to infection from a website $i$, when
the search provider implements an intervention controlled by the parameters
$\beta$ and $\sigma$. Recall that $\beta$ is the detection delay for infection
identification and $\sigma$ is the depreferencing parameter, i.e., the strength
of the response. The expectation of exposure to infection from website $i$ is then
\begin{align}
  \mathbb{E}[X_i(\beta,\sigma)] =& \sum_{x=0}^{\beta-1}\omega_i \frac{\rho \gamma
    (1-\gamma)^x}{\rho + \gamma} + \nonumber\\
  &\sum_{x=\beta}^\infty
  \omega_i\sigma^{x-\beta+1}\frac{\rho \gamma (1-\gamma)^x}{\rho +
    \gamma} \nonumber\\
  \label{eq:ExpectInfection}
  =& \frac{\omega_i \rho
\gamma}{\rho+\gamma}\left[\frac{1-(1-\gamma)^\beta}{\gamma} + \frac{\sigma(1-\gamma)^\beta}{1-(\sigma(1-\gamma))}\right]
\end{align}

\noindent The above expression simplifies to $\omega_iPr[S_i=I_i]$
when no intervention is taken, which would correspond to
$\beta=\infty$ or $\sigma=1$.

The other important random variable we are interested in is
$L_i(\beta,\sigma)$, which represents the traffic lost by a website
$i$ as a consequence of false positives. Following a similar analysis
to the earlier one for client exposure, if $F_x$ denotes being in the
false positive state for $x$ time steps, we have
\begin{equation}
  \label{eq:LossExact}
  Pr[S_i=F_x] = f Pr[S_i=U](1-(\gamma+\rho))^x
\end{equation}

\noindent The lost traffic at a specific time will be $\omega_i -
\hat{\omega}_i$. Substituting for $\hat{\omega}_i$ as given
by~\autoref{eq:depreferencing}, the expected traffic loss is
\begin{equation}
\begin{split}
  \label{eq:ExpectedLoss}
  &\mathbb{E}[L_i(\beta,\sigma)] = \\
  &\frac{\omega_i f \gamma (1-(\rho +
    \gamma))^\beta}{f + \gamma + \rho} \left[\frac{1}{\gamma+\rho} -
    \frac{\sigma}{1-\sigma(1-(\gamma+\rho))}\right]
\end{split}
\end{equation}

We can then define the {\em infection exposure}, which is the fraction of
traffic exposed to infection from all websites, as 
\begin{equation}
X(\beta,\sigma) = \frac{\sum_i^n X_i}{\sum_i^n \omega_i}
\end{equation}

and the overall {\em traffic loss} due to false positives as
\begin{equation}
L(\beta,\sigma) = \frac{\sum_i^n L_i}{\sum_i^n \omega_i}
\end{equation}

Using linearity of expectation, the expressions for
$\mathbb{E}[X(\beta,\sigma)]$ and $\mathbb{E}[L(\beta,\sigma)]$ are
simply those in \autoref{eq:InfectionExact} and \autoref{eq:LossExact}
respectively, while omitting $\omega_i$, specifically:

\begin{equation}
  \label{eq:ExpectInfectionRate}
  \begin{split}
    &\mathbb{E}[X(\beta,\sigma)] = \\
    &\frac{\rho
      \gamma}{\rho+\gamma}\left[\frac{1-(1-\gamma)^\beta}{\gamma} +
      \frac{\sigma(1-\gamma)^\beta}{1-(\sigma(1-\gamma))}\right]
  \end{split}
\end{equation}

\begin{equation}
\begin{split}
  \label{eq:ExpectedLossRate}
  &\mathbb{E}[L(\beta,\sigma)] = \\
  &\frac{f \gamma (1-(\rho +
    \gamma))^\beta}{f + \gamma + \rho} \left[\frac{1}{\gamma+\rho} -
    \frac{\sigma}{1-\sigma(1-(\gamma+\rho))}\right]
\end{split}
\end{equation}

We note that both of the infection exposure and the traffic loss are
independent of the distribution from which the $\omega_i$'s are drawn,
or how many servers there are.

The effectiveness of the depreferencing parameter, $\sigma$, and the 
detection delay, $\beta$, in the control strategy for
$\mathbb{E}[X(\beta,\sigma)]$, depends only on the recovery rate
$\gamma$. If $\gamma$ is particularly large (a fast recovery rate),
then any intervention will have a small effect. Only when websites are
slow to react to infections are interventions which alter traffic
likely to have significant impact.

Conversely, $\rho$ and $\gamma$ both affect
$\mathbb{E}[L(\beta,\sigma)]$. In particular, a decrease in the
infection rate $\rho$ or the recovery rate $\gamma$ will cause an
increase in loss due to false positives for a fixed false positive
rate $f$. Intuitively, a website that is unlikely to be in the
infected state is more vulnerable to being \emph{falsely infected}.

We now determine the variance in $X(\beta,\sigma)$ and
$L(\beta,\sigma)$. Because each of the $X_i$'s is independent and the
sum of the traffic is a constant,
\begin{equation}
Var[\frac{\sum_i^n
    X_i}{\sum_i^n\omega_i}]=\frac{\sum_i^nVar[X_i]}{\left(\sum_i^n\omega_i\right)^2}
\end{equation}

Additionally,
variance can be defined as $Var[X_i(\beta,\sigma)] =
\mathbb{E}[X_i(\beta,\sigma)^2] -
\mathbb{E}[X_i(\beta,\sigma)]^2$. Using these two facts and some
simple algebra we have:
\begin{equation}
  \label{eq:VarianceInfection}
  Var[X(\beta,\sigma)] =\left(\mathbb{E}[X(\beta,\sigma^2)] -
\mathbb{E}[X(\beta,\sigma)]^2\right) \frac{\sum_{i=1}^n
\omega_i^2}{(\sum_{i=1}^n \omega_i)^2}
\end{equation}

If the $\omega_i$'s are drawn from a distribution with finite variance
and expectation and $n$ is large, then we can apply the central limit
theorem to \autoref{eq:VarianceInfection} to rewrite it in terms of
the distribution of $\omega_i$'s
\begin{equation}
\begin{split}
  \label{eq:VarianceInfection2}
  &Var[X(\beta,\sigma)] = \\ &\left(\mathbb{E}[X(\beta,\sigma^2)] -
  \mathbb{E}[X(\beta,\sigma)^2]\right) \left(\frac{Var[\omega_i] +
    \mathbb{E}[\omega_i]^2}{n\mathbb{E}[\omega_i]^2}\right).
\end{split}
\end{equation}

Observe that \autoref{eq:VarianceInfection2} is monotonically
decreasing in the number of servers $n$. So as the population of
websites increases we expect the variance in the fraction of traffic
exposed to infection to go to 0.

It is almost certain, however, that the distribution of $\omega_i$ for
real webservers is heavy-tailed and does not have finite variance or
finite expectation~\cite{adamic00,clauset07,Meiss2010}. In the case of
a heavy-tailed or power-law distribution of $\omega_i$, the variance
$Var[X]$ does not converge to a single value for large $n$, but to a
distribution of values. Furthermore, because the sum of power-law
i.i.d. random variables exhibits heavy tailed behavior \cite{voit2005}
\cite{gnedenko1968}, the distribution of $Var[X(\beta,\sigma)]$ will
also exhibit heavy tailed behavior.

The sum of power law distributed variables can be approximated by the maximum
over the variables~\cite{zaliapin2005}, which means that the last fraction in
~\autoref{eq:VarianceInfection} can be approximated as $1$ for particularly
heavy tailed distributions and large $n$, i.e.
\begin{equation}
\frac{\sum_{i=1}^n \omega_i^2}{(\sum_{i=1}^n \omega_i)^2} \rightarrow 1
\end{equation}

If we take this as an upper bound, we see that improving either
$\sigma$ or $\beta$ to lower infection will also lower the variance in
the infection exposure rate. Depending on the value of the exponent in
the distribution of traffic, $Var[X(\beta,\sigma)]$ may not have
finite variance or expectation.  As we discuss later, this is
important because it implies that empirical studies of infection
exposure (or traffic loss) are likely to be highly sample dependent, and that
even significant changes to the variables like $\rho$ and $\gamma$ can
be hard to discern.

A similar analysis yields slightly different results for traffic loss:
\begin{equation}
\begin{split}
\label{eq:VarianceLoss}
  &Var[L(\beta,\sigma)]= \\ & \left(2\mathbb{E}[L(\beta,\sigma)] -
\mathbb{E}[L(\beta,\sigma^2)] - \mathbb{E}[L(\beta,\sigma)]^2\right)
\frac{\sum_{i=1}^n \omega_i^2}{\left(\sum_{i=1}^n
\omega_i\right)}
\end{split}
\end{equation}

\subsection{Critical Values}
\label{subsec:critvals}

In general, changing parameter values from one set, $(\beta,\sigma)$,
to another, $(\beta',\sigma')$, will result in a change in infection
exposure, i.e., $\mathbb{E}[X(\beta,\sigma)] \ne
\mathbb{E}[X(\beta',\sigma')]$. However, there could be some settings
of $\beta'$ and $\sigma'$, such that the outcome will not change,
i.e., $\mathbb{E}[X(\beta,\sigma)] =
\mathbb{E}[X(\beta',\sigma')]$. We call these settings, or transition
points, the {\em critical} values for the parameters.

The critical value, $\sigma_X$, for the depreferencing parameter is
the most important, because we expect that search providers will have
more control over $\sigma$ than $\beta$. For example, a new detection
algorithm may require a different $\beta'$; the search provider could
then use the critical value of $\sigma_X$ to ensure that the infection
exposure did not change.

To derive the critical value for the infection exposure, we first
calculate an expression for the precise value of $\sigma$ needed to
achieve a particular infection exposure rate
$\mathbb{E}[X(\beta,\sigma)] =\xi$, as
\begin{equation}
  \label{eq:SigmaHatInfection}
  \sigma = \frac{\frac{(\rho+\gamma)\xi}{\rho
\gamma}-\frac{1-(1-\gamma)^\beta}{\gamma}} {(1-\gamma)\left[\frac{(\rho
+ \gamma)\xi}{\rho \gamma} - \frac{1-(1-\gamma)^\beta}{\gamma}\right] +
  (1-\gamma)^\beta}
\end{equation}

We can then derive the critical value for the infection
exposure by substituting $\mathbb{E}[X(\beta',\sigma')]$ for $\xi$ in
~\autoref{eq:SigmaHatInfection}, which gives
\begin{equation}
  \label{eq:SigmaCritX}
  \sigma_X = \frac{a}{\gamma+a(1-\gamma)}
\end{equation}
\noindent where $a$ is defined as
\begin{equation}
  \label{eq:SigmaCritXa}
  a = 1- (1-\gamma)^{\beta-\beta'} + \frac{\sigma \gamma
    (1-\gamma)^{\beta-\beta'}}{1-\sigma(1-\gamma)}
\end{equation}

\autoref{eq:SigmaCritXa} shows the critical value needed to ensure the
infection exposure does not change when $\beta$ changes. An
alternative goal might be to ensure that the traffic loss due to false
positives does not change with a new value for $\beta$,
i.e. $\mathbb{E}[L(\beta',\sigma')] =
\mathbb{E}[L(\beta,\sigma)]$. This will be given by another critical
value, $\sigma_L$. Once again, we first derive an expression for the
precise value of $\sigma$ needed to attain a particular expected
traffic loss fraction $\mathbb{E}[L(\beta,\sigma)]=\lambda$,
\begin{equation}
  \label{eq:SigmaHatLoss}
  \sigma = \frac{\frac{1}{\rho + \gamma} -
    \frac{\lambda(f+\gamma+\rho)}{f\gamma(1-(\rho+\gamma))^\beta}}
  {1+(1-(\rho+\gamma))\left[\frac{1}{\rho
        + \gamma} -
      \frac{\lambda(f+\gamma+\rho)}{f\gamma(1-(\rho+\gamma))^\beta}\right]}
\end{equation}

Setting $\mathbb{E}[L(\beta',\sigma')] = \lambda$ in \autoref{eq:SigmaHatLoss},
we get
\begin{equation}
\label{eq:SigmaCritL}
\sigma_L = \frac{b}{1+b(1-\gamma-\rho)}
\end{equation}

\noindent where $b$ is defined as
\begin{equation}
  \label{eq:SigmaCritXb}
  b =\frac{1}{\gamma+\rho}
-(1-\gamma-\rho)^{\beta-\beta'}\left[\frac{1}{\gamma+\rho} -
\frac{\sigma}{1-\sigma(1-\gamma-\rho)}\right]
\end{equation}

As can be seen from \autoref{eq:SigmaCritL}, the critical value for the traffic
loss is independent of the false positive rate $f$.

Using~\autoref{eq:SigmaCritX} and ~\autoref{eq:SigmaCritL} in combination, a
search provider has the ability to decide how to adjust $\sigma$ to balance an
increase in the traffic loss against an increase in infection exposure.

\section{Experimental Results}
\label{sec:experiments}

To verify the results derived in Section~\ref{sec:analysis} we used a
Monte Carlo simulation of the model described in
Section~\ref{sec:model}. Unless otherwise noted, we used the following
parameter settings for all experiments: $\rho=0.01$, $\gamma=0.1$, and
$n=1000$. Although we believe that these parameter settings are
plausible, our goal is not to provide a precise match with real-world
outcomes, but rather to investigate more general consequences of
features such as variance and the comparative efficacy of
interventions. For each experiment, we conducted 1000 runs, and each
run was 75 time steps. This length is sufficient for the model to
reach a steady state.

We examine two different distributions throughout the experiments:
uniform, with $\omega_i \propto Uniform(0,1)$, and power law with
$\omega_i \propto x^{\alpha}$ with $\alpha=-1.4$. Although these two
distributions are likely not precisely representative of the real
world, they are useful in that they represent two possible
extremes of variance (finite and undefined).

\begin{figure}
 \centering
   \includegraphics[width=3.2in]{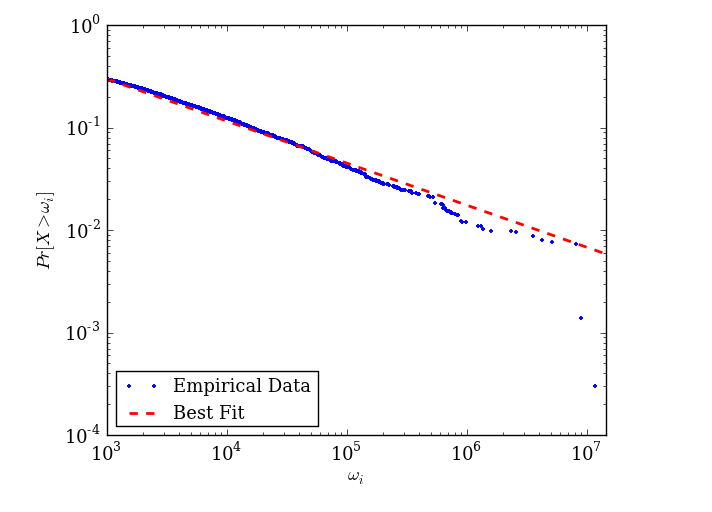}
   \caption{Empirically observed website traffic follows a power-law
     distribution with $\alpha=-1.4$.}
   \label{fig:Popularity}
\end{figure}

In reality, the distribution is likely heavy-tailed, possibly a
power-law~\cite{adamic00,clauset07,Meiss2010}. We found that a
power-law with an exponent of $\alpha=-1.4$ provides a good fit with
empirical data on website popularity, as can be seen in
\autoref{fig:Popularity}. We calculated the exponent for a random
sample of 10,000 websites listed in the top 1 million websites
according to the web-analytics firm Alexa, using estimates for the
daily number of visits obtained by querying the Alexa Web Information
Services API.\footnote{\url{http://aws.amazon.com/awis/}}

\subsection{Popularity Distribution}
\label{subsec:pop_dist}

According to the analysis in Section~\ref{sec:analysis}, distributions
of website popularity with undefined variance will result in large
fluctuations in client exposure to infection and will be highly
dependent on the sample of servers chosen. This is confirmed in our
experiments, as can be seen in
\autoref{fig:uniform_selection1000runs}. The uniform distribution of
website popularity results in low variance in client exposure
(\autoref{subfig:uniform_selection1000runs}), whereas the power law
website popularity results in very high variance, both in a single run
of the model and among different runs
(\autoref{subfig:pl_selection1000runs}).\footnote{Because the variance
  is undefined in general for a power-law, we substitute the run
  sample values of the $\omega_i$'s
  into~\autoref{eq:VarianceInfection} to compute the theoretical
  variance shown in~\autoref{subfig:pl_selection1000runs}.} For both
popularity distributions, the experimental average of the runs rapidly
converges to the expected steady-state value for $X$ (0.091), although
power-law distributions can yield $X$ values as high as 0.96 in
individual runs, an order of magnitude higher than the expected value.

\fig{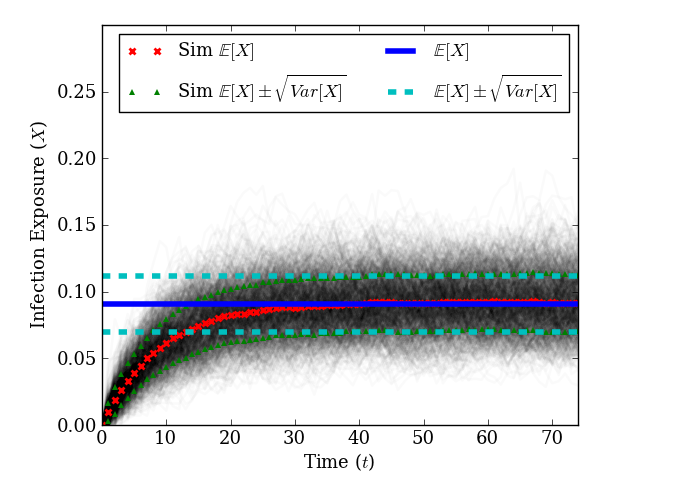}{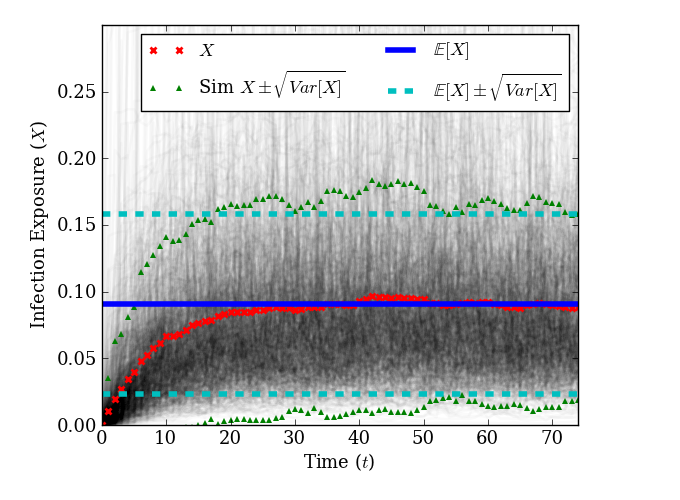}{Variation in
  client exposure to infection over time. Individual runs are light
  gray. {\em Sim} $X$ indicates the results of the simulation. Here
  $n=250$ to illustrate the effects of small sample sizes.}

\begin{figure*}[!t]
    \subfloat[Different runs with same parameters, $\rho=0.01$, $\gamma=0.1$.]{
      \includegraphics[width=3.4in]{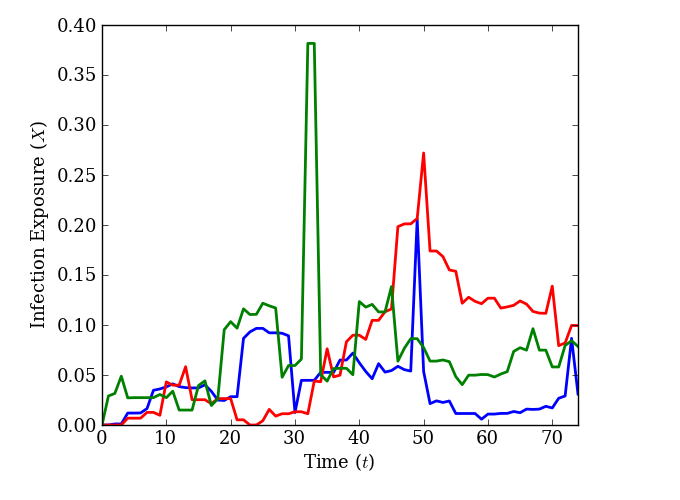}
     \label{subfig:pl_selection3runs}
   }
   \subfloat[Different runs with different parameters.]{
     \includegraphics[width=3.4in]{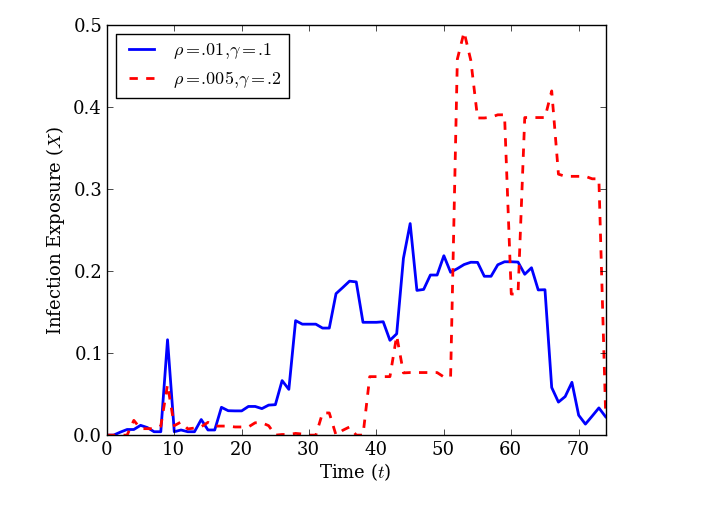}
     \label{subfig:pl_selection2rundiff}
   }
   \caption{Variation of infection exposure in individual runs for power-law
     distribution of website popularity.}
   \label{fig:pl_selection3runs}
\end{figure*}

\autoref{fig:pl_selection3runs} shows the variation in individual runs
more clearly. \autoref{subfig:pl_selection3runs} shows three different
runs of the simulation with the same parameters, $\rho=0.01,\gamma=0.1$.
There are large jumps in client exposure to infection that occur when
the more popular websites get infected, followed by plateaus before
those websites recover, and then abrupt drops after
recovery. \autoref{subfig:pl_selection2rundiff} shows two runs of the
model with different infection and recovery rate
parameters. Strikingly, the run with the infection rate cut in half
and the recovery rate doubled, seems to exhibit worse infection
behavior. This clearly illustrates why it might be difficult to
determine whether web security improvements are effective. The high
variance in the runs illustrates the importance of modeling, as
running experiments in the real world could require many trials over
long periods of time to reach conclusions with any confidence.

We also tested distributions other than uniform and power-law and confirmed the
theoretical prediction that distributions with finite variance produce low
variance in the measured outcome, whereas those with undefined variance produce
high variance in the measured outcome (results not shown).

\subsection{Interventions}
\label{subsec:interventions}

\autoref{fig:bl_timesuniform1000runs} demonstrates the effect of
varying the detection delay, $\beta$, on the steady state client
exposure rate. For both uniform and power-law popularity
distributions, blacklisting is effective only if implemented quickly,
i.e. before websites have had sufficient time to recover. The
likelihood of remaining infected for $t$ time steps is $(1-\gamma)^t$,
which becomes exponentially small for large $t$. For example, once
$\beta>40$, the steady state expected exposure is very close to the
theoretical value with no interventions (around 0.091). Thus, for
larger $\beta$, most infections will resolve before infected websites
are blacklisted. The precise relationship between $\gamma$ and $\beta$
is given by~\autoref{eq:ExpectInfection}.

\fig{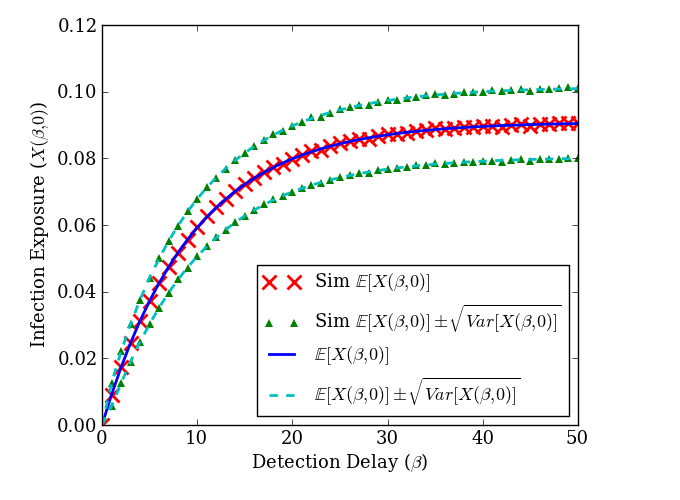}{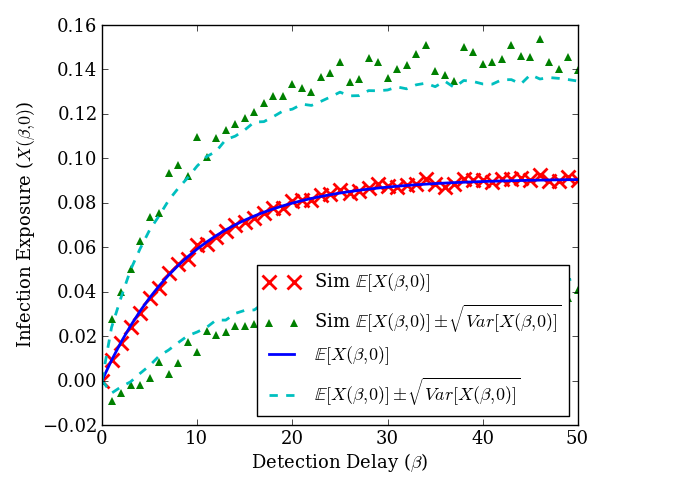}{Steady state client
  exposure to infection for various detection delays, $\beta$, with
  $\sigma=0$. }

The results of varying the depreferencing parameter, $\sigma$, are
shown in \autoref{fig:depreferencing1000runs_uniform}. Because
proportional depreferencing of popularity has an exponential impact on
the ranking (\autoref{eq:depreferencing}), even large values of
$\sigma$ can reduce infection rates significantly, for example, when
$\sigma=0.9$, the steady state client infection rate is half of the
baseline value.

\fig{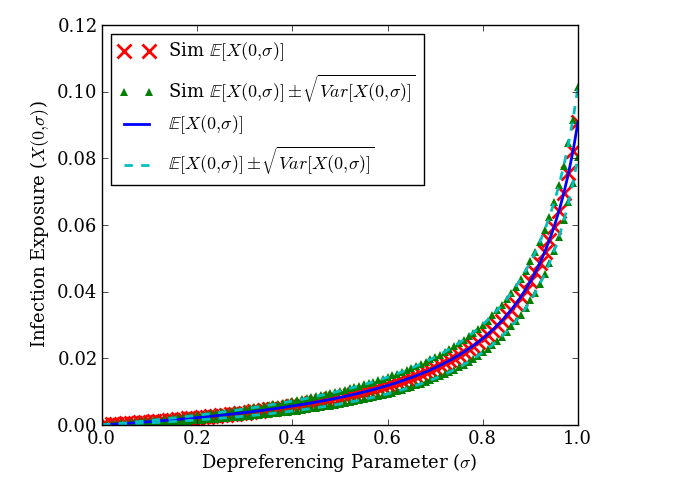}{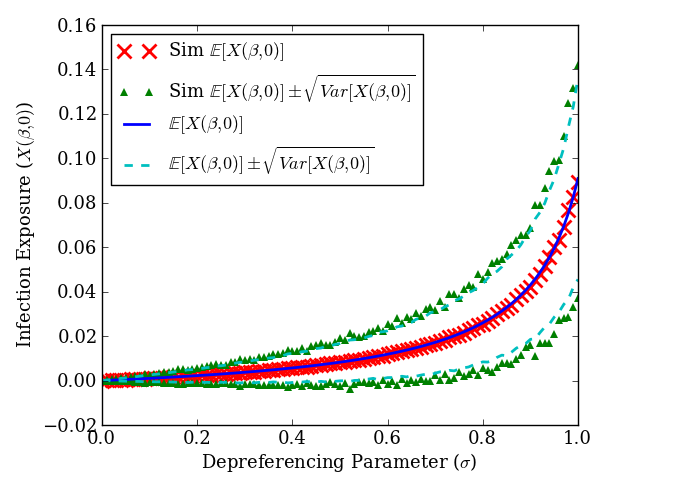}{Steady
  state client exposure to infection for various depreferencing
  adjustment values, $\sigma$, with $\beta=0$.}

Depreferencing gives finer control to search engines, because
adjusting $\sigma$ should be relatively easy, unlike trying to reduce
$\beta$, the control parameter for blacklisting. This finer control
might allow for algorithms that produce more false positives (which in
turn would reduce the number of missed infections), because the
effects of being mislabeled as infected could have far less impact on
a website that was moved down in the search rankings rather than being
blacklisted.

\subsection{False Positives}
\label{subsec:false_positives}

Depreferencing makes it feasible to use imprecise detection algorithms
that trade faster detection for higher false positives. In our model,
this would translate into a higher value for $f$, the false positive
probability. \autoref{fig:falsepositives1000runs_uniform} explores the
impact of $f$ on the change in traffic loss due to false
positives. Once again, a large variance in the website popularity
distribution has a large impact on the outcome, i.e. the traffic
loss. Further, as can be seen in
\autoref{fig:falsepositives1000runs_uniform}, reducing the false
positive rate is only worthwhile if it can be dropped below a certain
value (in this particular example, around $0.2$); when $f$ is high
enough, every website is mainly in the infected or falsely infected
state, and rarely in the uninfected state.

\fig{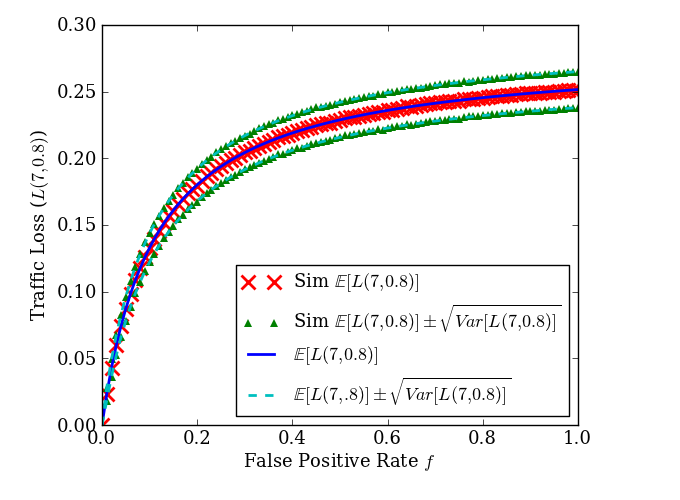}{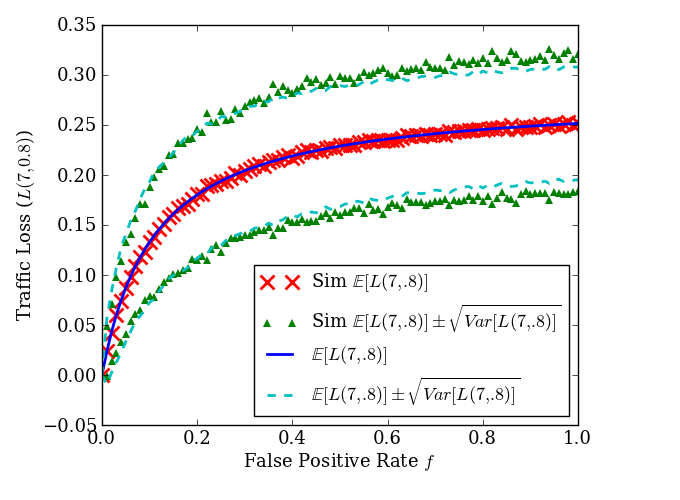}
    {Steady state normalized traffic loss for various false positive
      rates. Each data point is the average of 1000 runs. The value
      used for the depreferencing parameter was $\sigma=0.8$.}

\subsection{Exploring the Parameter Space}

\begin{figure*}[!t]
  \subfloat[Change in expected infection exposure]{
    \includegraphics[width=0.5\textwidth]{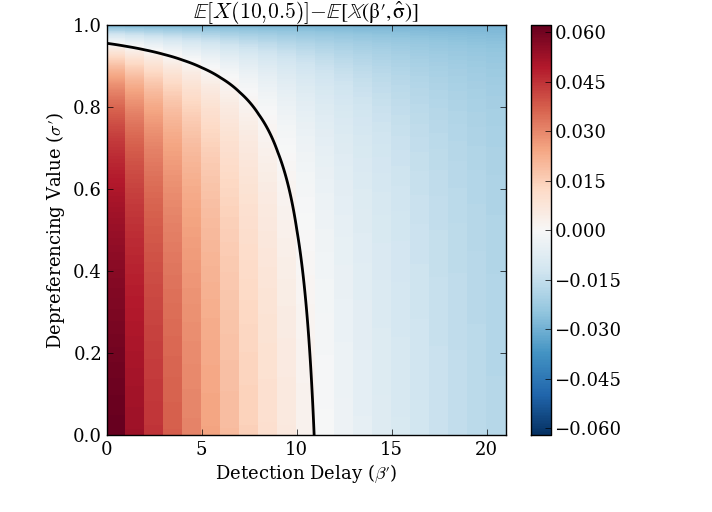}
    \label{subfig:heatmap_InfectionRateDifference}
  }
  \subfloat[Change in expected traffic loss]{
    \includegraphics[width=0.5\textwidth]{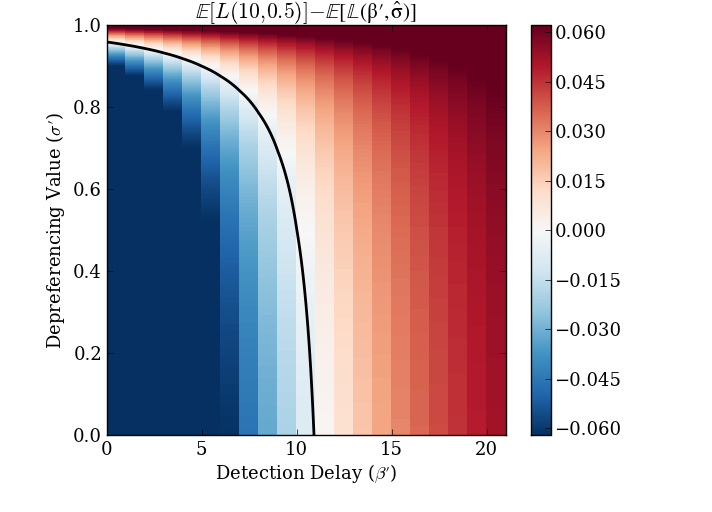}
    \label{subfig:heatmap_LossRateDifference}
  }
  \caption{Changes in outcomes when parameters $\beta'$ and $\sigma'$
    vary from a base of $\beta=10$ and $\sigma=0.5$. The solid lines correspond
    to the critical values, $\sigma'=\sigma_X$ in a) and $\sigma'=\sigma_L$ in b).}
  \label{fig:heatmap_changes}
\end{figure*}

\autoref{fig:heatmap_changes} shows how the expected infection
exposure and traffic loss change as the parameters $\sigma$ and
$\beta$ vary from a base setting of $\beta=10$ and $\sigma=0.5$. We
can see from the solid line at the critical value in
\autoref{subfig:heatmap_InfectionRateDifference} that changing the
depreferencing parameter, $\sigma$, can only correct for a small
increase in $\beta$, up to $\beta=11$. Beyond that, the expected
exposure increases, regardless of the setting of $\sigma$. The value
of $\sigma$ only starts to have a large positive impact if the
detection delay, $\beta$, drops significantly. We see similar results
for the change in expected traffic loss, as shown in
\autoref{subfig:heatmap_LossRateDifference}. Once again, only the
smallest increases in $\beta$ can be compensated for by increasing
$\sigma$. However, lack of compensation means a decrease in traffic
loss, which is a desirable outcome. We also see that it is easy to
adjust $\sigma$ to ensure that the traffic loss does not increase for
almost every change in $\beta$.

\begin{figure}[!t]
  \includegraphics[width=0.5\textwidth]{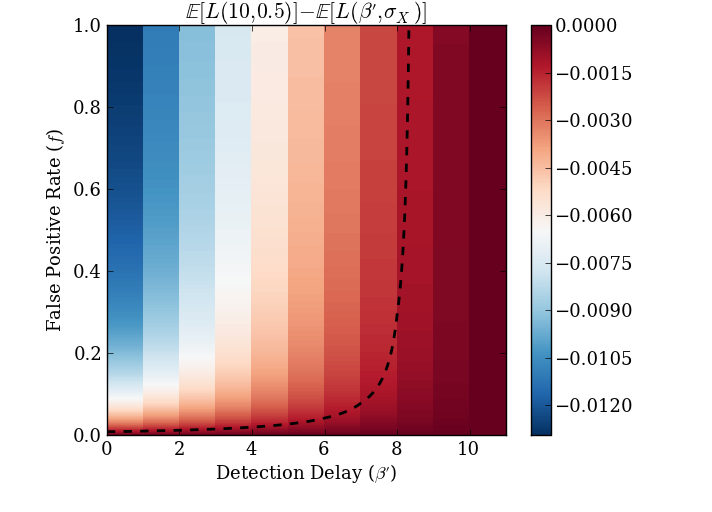}
  \caption{Change in expected traffic loss when expected infection exposure is
    kept constant, i.e. $\sigma'=\sigma_X$, the critical value. The base for
    comparison is $\beta=10$ and $\sigma=0.5$. The dotted line corresponds to a
    value of -0.1, i.e. an increase in traffic loss of 10\%.}
  \label{fig:trade_offs_critX_heatmap}
\end{figure}

It is clear that a faster response (reducing $\beta$) will reduce the
infection exposure rate, and any potential traffic loss can easily be
compensated for by changing $\sigma$. However, a faster response may
be less accurate and result in a higher false positive rate, $f$. We
explore this idea by again calculating the infection exposure with
base values $\beta=10$ and $\sigma=0.5$, and then calculating the
critical value $\sigma_X$ needed to maintain the same infection
exposure rate for a variety of $\beta'$ values. We then measure the
change in traffic loss
$\mathbb{E}[L(10,0.5)]-\mathbb{E}[L(\beta',\sigma_X)]$ for a variety
of false positive rates. The results can be see in
\autoref{fig:trade_offs_critX_heatmap}. Generally, a decrease in
detection delay, $\beta$, increases the traffic loss for a constant
false positive rate. If the false positive rate also goes up as
$\beta$ decreases, the problem is even worse. However, if the false
positive rate can be kept sufficiently small (below 0.1 in this
example), then there is flexibility to decrease the delay without
a major increase in traffic loss.

\section{Related Work}
\label{sec:related}

There are many approaches to combating web-based malware, including
the use of virtual machines or kernel extensions to check for
suspicious changes to the operating
system~\cite{DBLP:conf/ndss/MoshchukBGL06,DBLP:conf/ndss/WangBJRVCK06,Provos:Security08,BLADE:CCS2010},
emulating browsers to detect malicious
JavaScript~\cite{Cova:2010:DAD:1772690.1772720,Curtsinger:USENIX11},
and detecting campaigns that promote compromised sites to the top of
search results~\cite{John:USENIX11}. No technique is completely
effective at disrupting web-based malware, according to a study of
Google's data over more than four years~\cite{Rajab:TR11}. In our
view, one limiting factor is the choice of conservative approaches
that minimize false positives at the expense of speedy detection.  For
example, Provos et al.~\cite{Provos:Security08} choose to minimize
false positives in a system that allows explicit trade-offs between
false and true positives.

Depreferencing of search results is an example of a graduated
response, which is different from the binary, all-or-nothing, response
methods, such as blacklisting, that are usually taken in
cybersecurity. An early implementation of graduated response was a
Linux kernel extension called pH \cite{somayaji00}, which responded to
anomalous system call patterns by delaying subsequent system calls in
the offending process.  Other graduated responses operate by slowing
down, or throttling, outgoing requests
\cite{Williamson02a,hpimmunitymanager} in active networks
\cite{Hess03fidran:a}, Domain Name Service \cite{WongEtAl05a}, Border
Gateway Protocol \cite{KarlinEtAl06a}, and peer-to-peer networks
\cite{GhoshAndSchwartzbard99a}.  However, this is the first work we
are aware of that uses a graduated response outside of the time
domain.

Several studies have focused on alternative intervention strategies,
which could potentially be generalized using our depreferencing
method. For example, Hofmeyr et al. modeled responses available to
ISPs~\cite{Hofmeyr:WEIS11}. Other researchers have identified suitable
intervention strategies based on empirical research, which might also
be amenable to depreferencing. For example, Levchenko et
al.~\cite{Levchenko:SP11} found that criminals relied on just three
payment processors to collect money from victims, which led the
authors to recommend targeting the payment processors as a low-cost
intervention.  Similarly, Liu et al.~\cite{reginter:leet11}
empirically measured the effectiveness of pressuring registrars to
suspend spam-advertising domain names. In a related intervention,
Google has successfully pushed ad-filled sites down the results by
changes to its search-ranking algorithm~\cite{Moore:CCS11}, suggesting
that a similar effort to depreference malware-infected sites is
technically feasible.

\section{Discussion}
\label{sec:discussion}

A general theme of this research is the emphasis on modeling.
Modeling is a cost-effective way to explore intervention strategies,
including investigating novel ideas, without the expense of first
implementing them. As our results show, modeling can be particularly
helpful for understanding long-term trends in processes with high
variance, where direct experimentation can be misleading. Thoroughly
testing the interventions we explore in this paper would likely
require an unreasonable amount of time and money for any search
provider.

To the best of our knowledge, website depreferencing has not
previously been deployed to combat the drive-by-download problem. A
similar concept has been used previously in computer security
\cite{somayaji00,Balthrop2004}. Although we believe that
depreferencing is technically feasible, other issues may arise with
this type of response. For example, a policy that explicitly tolerates
false positives could trigger accusations of bias against search
engines.\footnote{The European Union is already investigating
  accusations that Google abused its power by preferring its own
  results over rivals. See
  \url{http://www.time.com/time/business/article/0,8599,2034138,00.html}.}
Another issue is how depreferencing might be gamed. For example, there
could be an incentive to deliberately infect competitors' websites, or
cause them to appear infected, so their search rankings are
demoted. Such industrial sabotage may in fact already happen.
However, the scope for it could increase if less precise,
false-positive tolerant detection mechanisms are used.

We have made several simplifying assumptions that we believe are
reasonable in the absence of more detailed information. For example,
we assume that website infection and client infection probabilities
are independent. In reality, this may not be the case. One variety of
drive-by-download malware steals the login credentials of users who
administer websites, enabling the malware to spread to those
websites. Hence, when a client is infected, the probability of
infecting one or more websites increases, corresponding to a change in
$\rho$. We have chosen not to model this form of malware spread,
because it has been observed only in a handful of outbreaks (e.g., one
Zeus variant in 2009~\cite{prevxftp}).  Another example is the
assumption that the distribution of website popularity is time
invariant, which is true in general, although the popularity of
individual websites can vary over
time~\cite{Krashakov:2006:URD:1167648.1167655}.  However, the
popularity of infected websites may change over time when attackers
attempt to promote compromised websites in search-engine
rankings~\cite{John:USENIX11}. In future, if sufficient information
can be attained, it may be possible to accurately model this
aspect. We believe, however, that even with more accurate information,
the heavy tailed nature of popularity will cause similar heavy tailed
behavior in infection exposure and traffic loss.

Another area of future work would be to focus on infections that spread in a
general network environment where a referral service (such as search) plays a
key role. Similar interventions could be applied
when infections are spread from website to website, rather than simply
exposing a client population. This could be a particularly good model
for controlling infections of malicious software in online social
networks.

In our analysis and modeling we disregard the effect of false
negatives, primarily because we assume that the response
methods we explore use the same detection mechanisms, subject to the
same false negative rates. Usually, in real detection systems,
reducing the accuracy of the system by increasing false positives
usually leads to a {\em decrease} in false negatives, a feature which
gives rise to the traditional ROC curve. We have insufficient data to
model this effect, but it suggests that the depreferencing mechanism
could have additional benefits beyond those shown by the model:
increasing tolerance of false positives could also improve the rate of
detection of compromised sites. 

Our focus in this research has been to develop a plausible model that
allows us to assess the impact of different interventions on the
spread of drive-by-download malware. Our goal is to show that modeling
can be a useful tool for search providers to use when considering
different interventions. We do not have access to data that could
enable us to make quantitative predictions about interventions. We
expect search providers to have much more relevant data, especially
information on the distribution of website popularity, the efficacy of
infected website detection and the recovery times for infection.

\section{Conclusion}
\label{sec:conclusion}

By building and analyzing plausible models, like the one presented in
this work, we are able to better understand where search providers and
web administrators should focus their efforts for reducing infections,
while avoiding the large-scale (and potentially expensive) experiments
needed to test interventions in the field. When there is a high
variance in the underlying distributions, such as the website
popularity, corresponding high variance in outcomes can make it
difficult to assess the comparative effectiveness of interventions in
one-off field experiments.

We proposed and explored a novel intervention strategy, called {\em
  depreferencing}, where a possibly infected website is moved down in
the search results, rather than outright blacklisted. Depreferencing
may be an attractive alternative to blacklisting for search providers
because it allows them to use less precise detection methods with
higher false positive rates, potentially increasing the speed of
response to infection and reducing the cost of detection. These
results imply great difficulty in determining empirically whether
certain website interventions are effective, and it suggests that
theoretical models such as the one described in this paper have an
important role to play in improving web security.

\section{Acknowledgments}
The authors gratefully acknowledge the support of DOE grant DE-AC02-05CH11231.
Stephanie Forrest acknowledges partial support of DARPA (P-1070-113237), NSF
(EF1038682,SHF0905236), and AFOSR (Fa9550-07-1-0532).

\bibliographystyle{IEEEtran}
\bibliography{IEEEabrv,paper}

\begin{thebibliography}{10}
\providecommand{\url}[1]{#1}
\csname url@samestyle\endcsname
\providecommand{\newblock}{\relax}
\providecommand{\bibinfo}[2]{#2}
\providecommand{\BIBentrySTDinterwordspacing}{\spaceskip=0pt\relax}
\providecommand{\BIBentryALTinterwordstretchfactor}{4}
\providecommand{\BIBentryALTinterwordspacing}{\spaceskip=\fontdimen2\font plus
\BIBentryALTinterwordstretchfactor\fontdimen3\font minus
  \fontdimen4\font\relax}
\providecommand{\BIBforeignlanguage}[2]{{%
\expandafter\ifx\csname l@#1\endcsname\relax
\typeout{** WARNING: IEEEtran.bst: No hyphenation pattern has been}%
\typeout{** loaded for the language `#1'. Using the pattern for}%
\typeout{** the default language instead.}%
\else
\language=\csname l@#1\endcsname
\fi
#2}}
\providecommand{\BIBdecl}{\relax}
\BIBdecl

\bibitem{Provos07}
N.~Provos, D.~McNamee, P.~Mavrommatis, K.~Wang, and N.~Modadugu, ``The ghost in
  the browser: Analysis of web-based malware,'' in \emph{Proc. 1st USENIX
  Workshop on Hot Topics in Understanding Botnets (HotBots'07)}, Cambridge, MA,
  Apr. 2007.

\bibitem{Provos:Security08}
N.~Provos, P.~Mavrommatis, M.~Rajab, and F.~Monrose, ``All your {iFrames} point
  to us,'' in \emph{Proc. 17th USENIX Security Symp.}, Aug. 2008.

\bibitem{GoogleSafe}
Google, ``Safe browsing api,'' \url{http://code.google.com/apis/safebrowsing/}.

\bibitem{Moore:CCS11}
T.~Moore, N.~Leontiadis, and N.~Christin, ``Fashion crimes: trending-term
  exploitation on the web,'' in \emph{Proc. ACM CCS'11}, Chicago, IL, Oct.
  2011.

\bibitem{zou2002}
C.~Zou, W.~Gong, and D.~Towsley, ``Code red worm propagation modeling and
  analysis,'' in \emph{Proceedings of the 9th ACM conference on Computer and
  communications security}.\hskip 1em plus 0.5em minus 0.4em\relax ACM, 2002,
  pp. 138--147.

\bibitem{Rajab:TR11}
M.~Rajab, L.~Ballard, N.~Jagpal, P.~Mavrommatis, D.~Nojiri, N.~Provos, and
  L.~Schmidt, ``Trends in circumventing web-malware detection,'' Google, Tech.
  Rep., Jul. 2011,
  \url{http://static.googleusercontent.com/external_content/untrusted_dlcp/res%
earch.google.com/en/us/archive/papers/rajab-2011a.pdf}.

\bibitem{GoogleCleanup}
U.~Parasites, ``Practical guide to dealing with google's malware warnings,''
  \url{http://www.unmaskparasites.com/malware-warning-guide/}.

\bibitem{adamic00}
L.~A. Adamic and B.~A. Huberman, ``Power-law distribution of the world wide
  web,'' \emph{Science}, vol. 287, p. 2115, 2000.

\bibitem{clauset07}
A.~Clauset, C.~Shalizi, and M.~Newman, ``Power-law distributions in empirical
  data,'' \emph{Arxiv preprint arxiv:0706.1062}, 2007.

\bibitem{Meiss2010}
M.~Meiss, B.~Gonçalves, J.~Ramasco, A.~Flammini, and F.~Menczer, ``Modeling
  traffic on the web graph,'' in \emph{Algorithms and Models for the
  Web-Graph}, ser. Lecture Notes in Computer Science, R.~Kumar and
  D.~Sivakumar, Eds.\hskip 1em plus 0.5em minus 0.4em\relax Springer Berlin /
  Heidelberg, 2010, vol. 6516, pp. 50--61.

\bibitem{voit2005}
J.~Voit, \emph{The statistical mechanics of financial markets}.\hskip 1em plus
  0.5em minus 0.4em\relax Springer Verlag, 2005.

\bibitem{gnedenko1968}
B.~Gnedenko, A.~Kolmogorov, K.~Chung, and J.~Doob, \emph{Limit distributions
  for sums of independent random variables}.\hskip 1em plus 0.5em minus
  0.4em\relax Addison-Wesley Reading, MA:, 1968, vol. 195.

\bibitem{zaliapin2005}
I.~Zaliapin, Y.~Kagan, and F.~Schoenberg, ``Approximating the distribution of
  pareto sums,'' \emph{Pure and Applied Geophysics}, vol. 162, no.~6, pp.
  1187--1228, 2005.

\bibitem{DBLP:conf/ndss/MoshchukBGL06}
A.~Moshchuk, T.~Bragin, S.~D. Gribble, and H.~M. Levy, ``A crawler-based study
  of spyware in the web,'' in \emph{NDSS}, 2006.

\bibitem{DBLP:conf/ndss/WangBJRVCK06}
Y.-M. Wang, D.~Beck, X.~Jiang, R.~Roussev, C.~Verbowski, S.~Chen, and S.~T.
  King, ``Automated web patrol with strider honeymonkeys: Finding web sites
  that exploit browser vulnerabilities,'' in \emph{NDSS}, 2006.

\bibitem{BLADE:CCS2010}
L.~Lu, V.~Yegneswaran, P.~Porras, and W.~Lee, ``Blade: An attack-agnostic
  approach for preventing drive-by malware infection,'' \emph{Proceedings of
  the 17th ACM Conference on Computer and Communications Security}, 2010.

\bibitem{Cova:2010:DAD:1772690.1772720}
\BIBentryALTinterwordspacing
M.~Cova, C.~Kruegel, and G.~Vigna, ``Detection and analysis of
  drive-by-download attacks and malicious javascript code,'' in \emph{Proc. WWW
  '10}, 2010, pp. 281--290. [Online]. Available:
  \url{http://doi.acm.org/10.1145/1772690.1772720}
\BIBentrySTDinterwordspacing

\bibitem{Curtsinger:USENIX11}
C.~Curtsinger, B.~Livshits, B.~Zorgn, and C.~Seifert, ``{ZOZZLE}: Fast and
  precise in-browser javascript malware detection,'' in \emph{Proc. 20th USENIX
  Security Symp.}, Aug. 2011.

\bibitem{John:USENIX11}
J.~John, F.~Yu, Y.~Xie, M.~Abadi, and A.~Krishnamurthy, ``{deSEO}: Combating
  search-result poisoning,'' in \emph{Proc. USENIX Security Symp. 2011}, San
  Francisco, CA, 2011.

\bibitem{somayaji00}
A.~Somayaji and S.~Forrest, ``Automated response using system-call delays,'' in
  \emph{In Proceedings of the 9th USENIX Security Symposium}, 2000, pp.
  185--197.

\bibitem{Williamson02a}
M.~M. Williamson, ``Throttling viruses: Restricting propagation to defeat
  malicous mobile code,'' in \emph{Proc. ACSAC '02}, Las Vegas, Nevada, Dec.
  2002.

\bibitem{hpimmunitymanager}
HP, ``Immunity manager,'' Website, http://www.hp.com/rnd/
  pdfs/ProCurve\_Network\_Immunity\_Manager1\_0.pdf.

\bibitem{Hess03fidran:a}
A.~Hess, M.~Jung, and G.~Schfer, ``Fidran: A flexible intrusion detection and
  response framework for active networks,'' in \emph{ISCC '03}, 2003.

\bibitem{WongEtAl05a}
C.~Wong, S.~Bielski, A.~Studer, and C.~Wang, ``On the effectiveness of rate
  limiting mechanisms,'' in \emph{Proc. RAID '05}, 2005.

\bibitem{KarlinEtAl06a}
J.~Karlin, J.~Rexford, and S.~Forrest, ``Pretty good bgp: Improving bgp by
  cautiously adopting routes,'' in \emph{Proc. CNP '06}, 2006.

\bibitem{GhoshAndSchwartzbard99a}
A.~Ghosh and A.~Schwartzbard, ``A study in using neural networks for anomaly
  and misuse detection,'' in \emph{Proc. USENIX Security Symp.}, 1999.

\bibitem{Hofmeyr:WEIS11}
S.~Hofmeyr, T.~Moore, B.~Edwards, S.~Forrest, and G.~Stelle, ``Modeling
  {I}nternet-scale policies for cleaning up malware,'' in \emph{Proc. 10th
  Workshop on the Economics of Information Security}, 2011.

\bibitem{Levchenko:SP11}
K.~Levchenko, N.~Chachra, B.~Enright, M.~Felegyhazi, C.~Grier, T.~Halvorson,
  C.~Kanich, C.~Kreibich, H.~Liu, D.~McCoy, A.~Pitsillidis, N.~Weaver,
  V.~Paxson, G.~Voelker, and S.~Savage, ``Click trajectories: End-to-end
  analysis of the spam value chain,'' in \emph{Proc. IEEE Sym. and Security and
  Privacy}, Oakland, CA, 2011.

\bibitem{reginter:leet11}
H.~Liu, K.~Levchenko, M.~F{\'e}legyh{\'a}zi, C.~Kreibich, G.~Maier, G.~M.
  Voelker, and S.~Savage, ``On the effects of registrar-level intervention,''
  in \emph{{Proc. USENIX Workshop on Large-scale Exploits and Emergent Threats
  (LEET)}}, Boston, MA, March 2011.

\bibitem{Balthrop2004}
\BIBentryALTinterwordspacing
J.~Balthrop, S.~Forrest, M.~E.~J. Newman, and M.~M. Williamson, ``Technological
  networks and the spread of computer viruses,'' \emph{Science}, vol. 304, no.
  5670, pp. 527--529, 2004. [Online]. Available:
  \url{http://www.sciencemag.org/content/304/5670/527.short}
\BIBentrySTDinterwordspacing

\bibitem{prevxftp}
J.~Erasmus, ``Compromised ftp details being exploited by in the wild malware,''
  Jun. 2009,
  \url{http://www.prevx.com/blog/132/Compromised-FTP-details-being-exploited-b%
y-in-the-wild-malware.html}.

\bibitem{Krashakov:2006:URD:1167648.1167655}
\BIBentryALTinterwordspacing
S.~A. Krashakov, A.~B. Teslyuk, and L.~N. Shchur, ``On the universality of rank
  distributions of website popularity,'' \emph{Comput. Netw.}, vol.~50, pp.
  1769--1780, August 2006. [Online]. Available:
  \url{http://portal.acm.org/citation.cfm?id=1167648.1167655}
\BIBentrySTDinterwordspacing

\end{thebibliography}
\end{document}